\documentclass[conference]{IEEEtran}
\IEEEoverridecommandlockouts

\usepackage{cite}
\usepackage{url}
\usepackage{amsmath,amssymb,amsfonts}

\usepackage{graphicx}
\usepackage{caption}
\usepackage{subcaption}
\usepackage{multirow}
\usepackage{algorithm}
\usepackage{algpseudocode}
\usepackage{comment}
\usepackage{textcomp}
\usepackage{xcolor}
\def\BibTeX{{\rm B\kern-.05em{\sc i\kern-.025em b}\kern-.08em
    T\kern-.1667em\lower.7ex\hbox{E}\kern-.125emX}}

\def\projectname{{STAR-FL}}

\begin{document}

\title{STAR-FL: Secure Federated Learning with Spatial-Temporal Analysis and Robust Aggregation}

\author{\IEEEauthorblockN{Nawrin Tabassum}
\IEEEauthorblockA{\textit{Department of Computer Science and Engineering} \\
\textit{Ahsanullah University of Science and Technology}\\
Dhaka, Bangladesh \\
nawrin.cse@aust.edu}
\and
\IEEEauthorblockN{Yanzhao Wu}
\IEEEauthorblockA{\textit{Knight Foundation School of Computing and Information Sciences} \\
\textit{Florida International University}\\
Miami, FL, USA \\
yawu@fiu.edu}
}

\maketitle

\begin{abstract}
Data poisoning attacks pose serious security threats to Federated Learning (FL) systems in Computer Vision. Despite growing research attention, two key challenges remain for existing defense techniques: (1) accurately distinguishing between benign and malicious model updates and (2) effectively mitigating the influence of poisoned model updates during model aggregation. To address these challenges, we propose a novel defense framework against targeted poisoning attacks with Spatial-Temporal Analysis and Robust aggregation for FL (STAR-FL). First, we employ spatial-temporal clustering to identify and remove potentially malicious updates from the FL training process. Second, we adjust the learning rate during aggregation to mitigate the impact of any malicious updates that evade detection. Third, we conduct extensive experiments across multiple benchmark datasets to evaluate the spatial-temporal analysis and robust aggregation in STAR-FL. Experimental results demonstrate their synergistic effect in enabling STAR-FL to effectively protect FL and consistently outperform state-of-the-art defenses against targeted poisoning attacks, significantly reducing Attack Success Rates (ASRs). The source code is available at \url{https://github.com/mlsysx/STAR-FL}.
\end{abstract}

\begin{IEEEkeywords}
Federated Learning, Targeted Poisoning Attack, Backdoor Defense, Spatial-Temporal Analysis, Robust Aggregation.
\end{IEEEkeywords}

\section{Introduction}
Federated Learning (FL)~\cite{mcmahan2017communication} is a distributed Machine Learning (ML) paradigm that is widely used for privacy-preserving model training. FL enables decentralized training across multiple clients, where each client owns a local dataset. At each communication round, a subset of clients is selected to participate in the FL training. The selected client receives the current global model, performs local training on its private data, and returns model updates to the central server for aggregation. Despite several advantages, FL remains vulnerable to a wide range of adversarial attacks~\cite{wang2020attack,cao2022mpaf,xie2019dba,tolpegin2020data,das2023privacy,sun2025partner,wei2023model,tabassum2024efficiency}. These attacks not only compromise the privacy of client data but also degrade model performance by disrupting the learning process. In data poisoning attacks, malicious clients modify their local data samples and send poisoned updates to the server. During aggregation, these poisoned updates are combined with the benign updates from the honest clients. In untargeted poisoning attacks~\cite{cao2022mpaf,fang2020local}, adversaries aim to degrade the global model performance by sending poisoned updates. These updates deviate from the correct optimization direction and cause the global model to misclassify input samples. Consequently, the model accuracy decreases as the server aggregates both honest and poisoned updates. 
In contrast, targeted poisoning attacks~\cite{wang2020attack,bagdasaryan2020backdoor,zhang2022neurotoxin} cause the model to misclassify specific inputs while maintaining high accuracy on benign data, making these attacks particularly difficult to detect. 

Several defenses have been proposed to mitigate these adversarial threats in FL. One line of work focuses on pre-aggregation defenses, which aim to identify and exclude malicious client updates before aggregation. These approaches typically leverage unsupervised learning techniques to distinguish between benign and malicious updates by analyzing spatial characteristics, such as gradient similarities or distance measures~\cite{rieger2022deepsight,nguyen2022flame,fung2018mitigating}. Some methods detect malicious clients based on temporal behaviors and exclude their updates from the aggregation process~\cite{zhang2022fldetector}. Another line of work employs robust aggregation techniques~\cite{pillutla2022robust,ozdayi2021defending,yin2018byzantine}, which attempt to mitigate the influence of adversarial updates by reweighting client contributions during aggregation. Although these defenses can improve robustness, they often mistakenly discard benign updates, resulting in the loss of useful information and degraded model performance. These limitations highlight the pressing need for a holistic defense framework that jointly performs reliable attack detection and robust aggregation.

In this study, we propose \projectname{}, a defense framework that integrates spatial-temporal analysis with robust aggregation to secure FL systems against targeted poisoning attacks. We specifically focus on backdoor attacks where malicious clients implant a trigger pattern into specific training samples and alter their labels. Our key insight is that malicious client updates often exhibit abnormal spatial relationships with benign updates and temporal inconsistencies with their own historical behavior. Leveraging this observation, \projectname{} jointly analyzes the spatial similarity of client updates and their temporal consistency across training rounds to identify suspicious behaviors. Updates flagged as malicious are excluded prior to aggregation. 
To further mitigate the influence of adversarial updates that evade spatial-temporal detection, \projectname{} incorporates a robust aggregation strategy that adaptively adjusts the server learning rate to reweight client contributions during each round. This mechanism preserves effective learning on the main task while suppressing the impact of poisoned updates. By combining spatial-temporal analysis with robust aggregation, \projectname{} provides a unified defense against backdoor attacks in FL. 
We evaluate \projectname{} on three benchmark datasets: Fashion-MNIST (FMNIST)~\cite{xiao2017fashion}, CIFAR-10~\cite{krizhevsky2009learning}, and CIFAR-100~\cite{krizhevsky2009learning} under multiple poisoning attacks, including Backdoor Attack, Model Replacement Attack, Distributed Backdoor Attack, and Adaptive Attack. Experimental results show that \projectname{} significantly reduces attack success rates while maintaining high model utility and outperforms state-of-the-art defenses.
The main contributions of our work can be summarized as follows:

\begin{itemize}
    \item We propose \projectname{}, a novel defense framework that effectively detects malicious client updates and mitigates backdoor attacks in federated learning. 
    \item We design a spatial-temporal analysis mechanism that examines the similarity among client updates and their temporal consistency across rounds to identify poisoned updates. We further integrate a robust aggregation strategy that suppresses the influence of malicious updates while preserving the contributions of benign clients.
    \item We conduct extensive experiments on multiple benchmark datasets and attack scenarios, demonstrating that STAR-FL significantly reduces attack success rates while maintaining high model utility.
\end{itemize}

\section{Related Work}
In targeted poisoning attacks, adversaries manipulate local training data to implant hidden behaviors into the global model. A common strategy is to insert trigger patterns into training samples and associate them with a target label to inject a backdoor, which causes the model to misclassify triggered inputs while maintaining high accuracy on clean data~\cite{fang2023vulnerability,fung2018mitigating,zhao2024backdoor}. Typically, attackers inject the same trigger pattern into targeted inputs~\cite{gu2017badnets}. Another attack variant is the distributed backdoor attack (DBA)~\cite{xie2019dba}, which distributes trigger patterns across multiple malicious clients to evade detection. Label-flipping attacks~\cite{tolpegin2020data,jha2023labelflip} modify the labels of samples from a source class to a target class, thereby poisoning the learning process without introducing explicit triggers.

Various defense mechanisms have been proposed to mitigate the impact of poisoned updates in federated learning~\cite{munoz2019byzantine,wu2020mitigating}. One line of work focuses on \textit{update filtering}, where malicious updates are detected and excluded before aggregation. Several studies cluster model updates based on distance measures, such as cosine distance, to separate benign and malicious clients~\cite{nguyen2022flame,rieger2022deepsight}. Dimensionality reduction techniques, such as PCA, have been investigated to improve clustering performance~\cite{tolpegin2020data}. In~\cite{zhang2022fldetector}, a client is flagged as malicious if its updates are inconsistent with those predicted by the server based on its historical behavior. However, these defense approaches often tend to misclassify and discard benign updates, harming global model utility. Robust aggregation techniques~\cite{pillutla2022robust,yin2018byzantine} have also been studied to reduce the influence of adversarial clients during FL training. Robust Learning Rate (RLR)~\cite{ozdayi2021defending} adjusts the server learning rate based on the signs of client updates to suppress adversarial contributions. Nevertheless, such methods are typically effective only when the number of malicious clients is very small, and residual poisoned updates may still degrade global model performance.

\section{Problem Statement}\label{AA}

\subsection{Threat Model}

The goal of the attack is to manipulate the model in such a way that it correctly classifies benign samples. However, poisoned samples are misclassified into the attacker-specified target class, resulting in a high attack success rate. The attacker aims to ensure that the global model achieves high accuracy on both the main task and the backdoor task. The attackers do not manipulate the global model directly. Rather, they exploit the local datasets by inserting a trigger pattern into specific samples and altering their labels. We assume that the attackers have partial knowledge about the FL system, i.e., local training data. However, they have no knowledge about the aggregator or central server.

Let a model $f$ be trained on $(x,y)$ where $x$ is the input sample and $y$ is the ground truth label. The trained model predicts the output $f(x)$ for clean input $x$. When the attacker poisons the input $x$ by adding some trigger $\mu$, the model misclassifies the poisoned sample $x+\mu$ into a specific target label $y'$ such that $f(x+\mu) = y'$. We consider a backdoor attack where the attacker's goal is to retain the model's behavior on benign data while modifying its behavior on poisoned data in the presence of a trigger. The following objective is optimized by the attacker. 
\begin{equation}
\theta^* = \min_{\theta} \sum_{i \in D_H} L_1(x_i, y_i) + \sum_{i \in D_P} L_2(x_i + \mu, y_i')
\end{equation} 
where the loss functions $L_1$ and $L_2$ are simultaneously minimized for clean dataset $D_H$ and poisoned dataset $D_P$.

\begin{figure}
\centering
\begin{subfigure}{0.1\textwidth}
    \includegraphics[width=\textwidth]{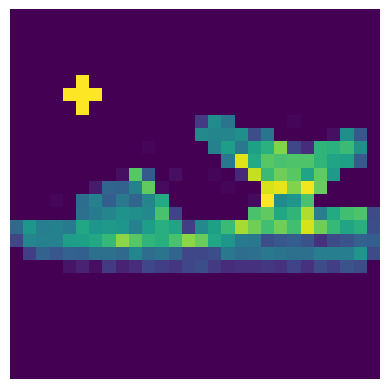}
    \caption{}
    \label{fig:first}
\end{subfigure}
\begin{subfigure}{0.1\textwidth}
    \includegraphics[width=\textwidth]{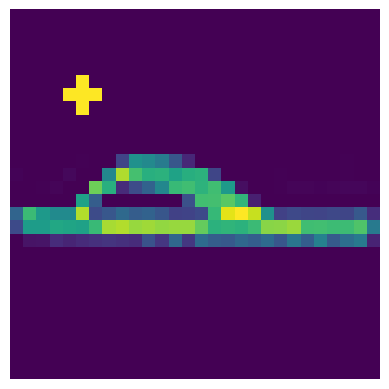}
    \caption{}
    \label{fig:second}
\end{subfigure}
\begin{subfigure}{0.1\textwidth}
    \includegraphics[width=\textwidth]{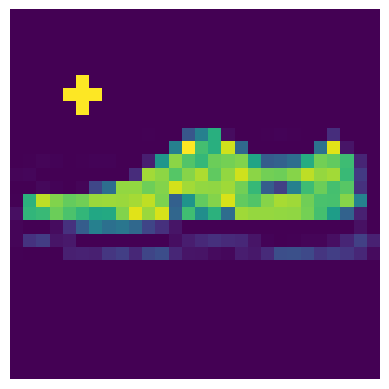}
    \caption{}
    \label{fig:third}
\end{subfigure} 

\begin{subfigure}{0.1\textwidth}
    \includegraphics[width=\textwidth]{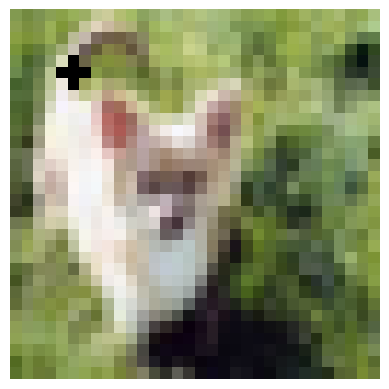}
    \caption{}
    \label{fig:first1}
\end{subfigure}
\begin{subfigure}{0.1\textwidth}
    \includegraphics[width=\textwidth]{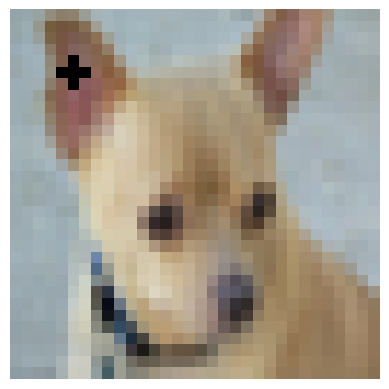}
    \caption{}
    \label{fig:second1}
\end{subfigure}
\begin{subfigure}{0.1\textwidth}
    \includegraphics[width=\textwidth]{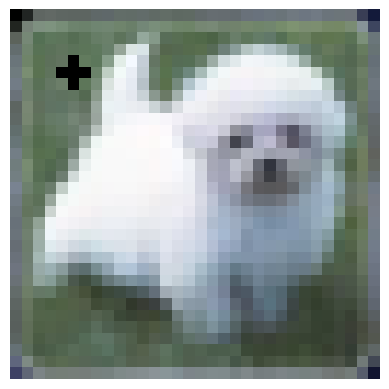}
    \caption{}
    \label{fig:third1}
\end{subfigure}  

\caption{Poisoned image examples from FMNIST (top) and CIFAR-10 (bottom) datasets. A plus pattern is inserted as a trigger in the top left of the images.}
\label{fig:figures}
\vspace{-5mm}
\end{figure}

\subsection{Defense Goals}

Let $\mathcal{M}_t \subseteq \mathcal{K}_t$ be the set of malicious clients at round $t$, and \( \mathcal{D}_{H} \) and \( \mathcal{D}_{P} \) represent the honest and poisoned data distributions respectively. Given a global model \( f_W \) with weights \( W \) and a loss function \( \ell \), the defense goal is to optimize the objective defined in Eq.~\eqref{eq4} by minimizing the main task loss while maximizing the backdoor task loss. 
\begin{equation}
\begin{split}
\mathcal{L}_{W} = \min_{W} \; \mathbb{E}_{(x, y) \sim \mathcal{D}_{H}}[\ell(f_W(x), y)] &+ \\
\max_{W} \; \mathbb{E}_{(x, y) \sim \mathcal{D}_{P}}[\ell(f_W(x), y)]
\end{split}
\label{eq4}
\end{equation}

Our goal is to design an effective defense method in which the global model would demonstrate good performance on the main task by accurately classifying benign samples. At the same time, when attackers are present, the model would not learn the backdoor task or misclassify poisoned samples into the target label. It would correctly classify both benign and poisoned samples into their true labels.

\begin{algorithm}[!h]
\caption{STAR-FL Framework}
\label{algo:starfl}
\footnotesize
\begin{algorithmic}[1]
\Statex \textbf{Input:} Number of rounds $T$, number of clients $N$, server learning rate $\eta$, decay parameter $\beta$, sign threshold $\delta$, local datasets $D_i$ for each client $i$. 
\Statex \textbf{Output:} Final global model $W_G$. 
\Statex \textbf{Initialize:} Global model $W_0$, historical updates $R^0_i = 0$ for each client $i$.

\For{each round $t = 1$ to $T$}
    \For{each client $i = 1$ to $N$}
        \State Receive global model $W_{t-1}$
        \State Perform local training and send $\Delta W^i_t$
    \EndFor
\State // Spatial-Temporal Analysis
    \State $\{h_t\} \gets \Call{DetectMaliciousClients}{\{\Delta W^i_t\}, \{R^{t-1}_i\}, \beta}$

\State // Robust Aggregation
    \For{each dimension $k \in \{1,2,3,....,K\}$}
        \State Compute sum of signs:
        \[
        M_{t_k} = \sum_{i \in h_t} sign\;(\Delta W^i_{t_k})
        \]
        \State Adjust learning rate:
        \[
        \eta^t_k = 
        \begin{cases}
        \eta, & if\; |M_{t_k}| > \delta \\
        -\eta, & otherwise
        \end{cases}
        \]
    \EndFor

    \State Update global model:
    \[
    W_t = W_{t-1} + \sum_{i \in h_t} \frac{|D_i|}{\sum_{j \in h_t} |D_j|} \cdot (\eta^t \odot \Delta W^i_t)
    \]
\EndFor
\State \Return Final model $W_G$
\end{algorithmic}
\end{algorithm}

\section{Overview of STAR-FL}

Our proposed defense framework \projectname{} consists of the following steps. First, we take the final layer model updates and reduce their dimensionality. Then we separate these updates into two clusters. Second, we predict each client's updates in a specific round using their historical updates. Then we compare the predicted updates with the actual updates received from the clients after local training. Updates that are identified as malicious by both spatial and temporal analyses are removed from that round. Third, we aggregate the remaining model updates while adjusting the server learning rate based on the sum of update signs. Algorithms \ref{algo:starfl} and \ref{algo:detect} demonstrate the detailed steps of \projectname{}. Our work is conducted entirely within the Horizontal Federated Learning (HFL) setting, where each client holds a distinct subset of the dataset. Our problem formulation, threat model, attack, and defense are all implemented within the HFL setting.  

\begin{algorithm}[!h]
\caption{DetectMaliciousClients}
\label{algo:detect}
\footnotesize
\begin{algorithmic}[1]
\Statex \textbf{Input:} Local model updates $\{\Delta W^i_t\}$, historical updates $\{R^{t-1}_i\}$, decay parameter $\beta$. 
\Statex \textbf{Output:} Honest clients $\{h_t\}$.
\State // Spatial Analysis
\For{each client $i = 1$ to $N$}
    \State Extract last layer gradients $\nabla W^i_t$
    \State Apply PCA to obtain dimensionality reduced gradient $V^t_i$
\EndFor
\State Apply K-means clustering on $\{V^t_i\}$ to obtain clusters $C_1, C_2$
\State Compute average cosine similarity $S_{C_p}$ for each cluster:
\[
S_{C_p} = avg \; { \frac{V_i \cdot V_j}{\|V_i\| \, \|V_j\|}; } \qquad \forall \; {i,j \in C_p, i \neq j}
\]

\If{$S_{C_1} > S_{C_2}$}
    \State Clients in $C_1$ are potentially malicious
\Else
    \State Clients in $C_2$ are potentially malicious
\EndIf
\State // Temporal Analysis
\For{each client $i = 1$ to $N$}
    \State Predict update:
    \[
    R^t_i = \beta R^{t-1}_i + (1 - \beta) \Delta W_t
    \]
    \State Compute similarity:
    \[
    \gamma^t_i = \frac{R^t_i \cdot \Delta W_t^i}{\|R^t_i\| \,\|\Delta W_t^i\|}
    \]
\EndFor
\State Apply KDE on $\{\gamma^t_i\}$ to detect temporal outliers
\State $\{h_t\} \gets$ clients detected as honest in both analyses
\State \Return $\{h_t\}$
\end{algorithmic}
\end{algorithm}

\subsection{Spatial Analysis}

\begin{figure}
\centering
\begin{subfigure}{0.49\linewidth}
    \includegraphics[width=\linewidth]{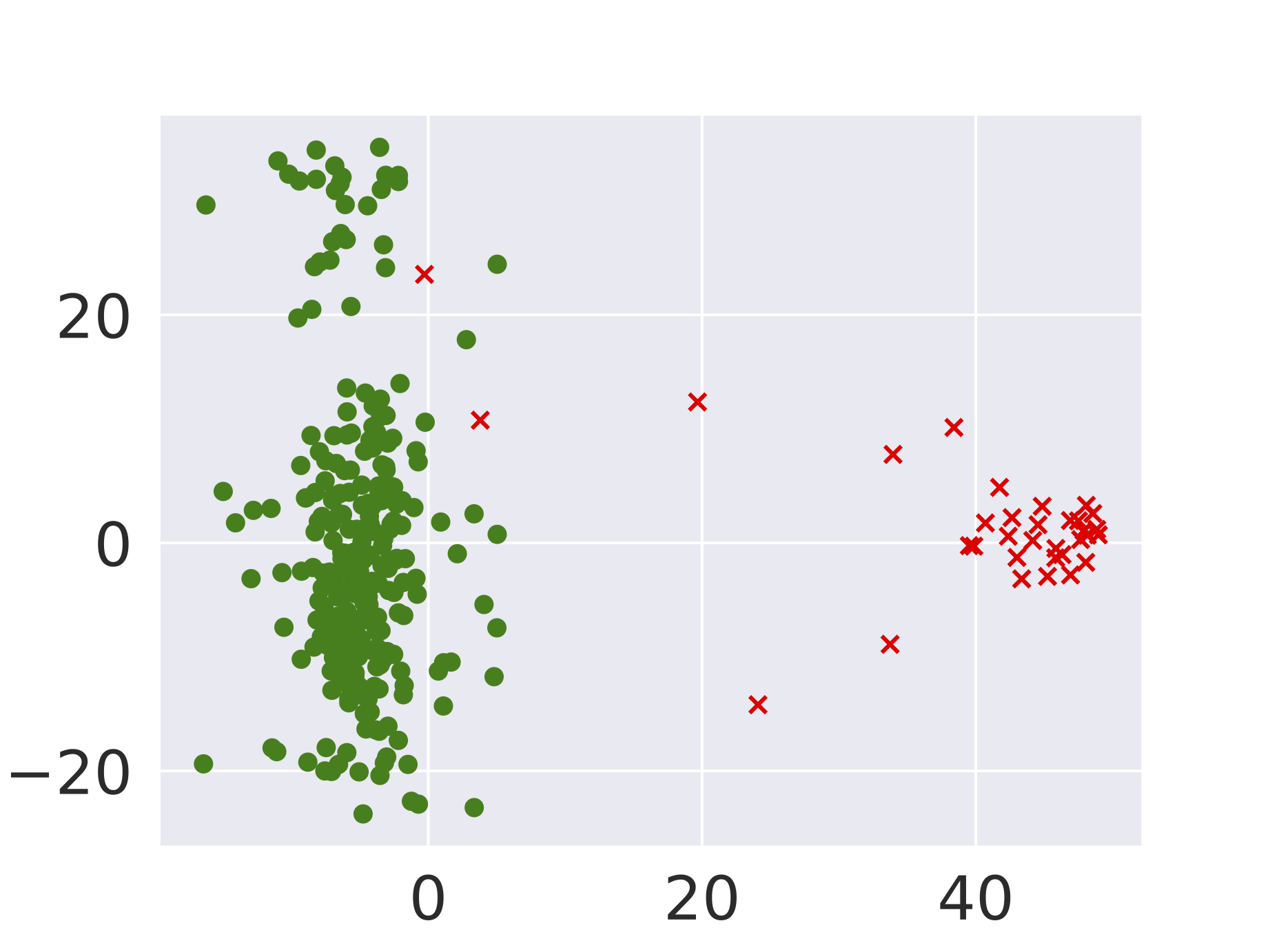}
    \caption{FMNIST}
    \label{fig:first2}
\end{subfigure}
\hfill
\begin{subfigure}{0.49\linewidth}
    \includegraphics[width=\linewidth]{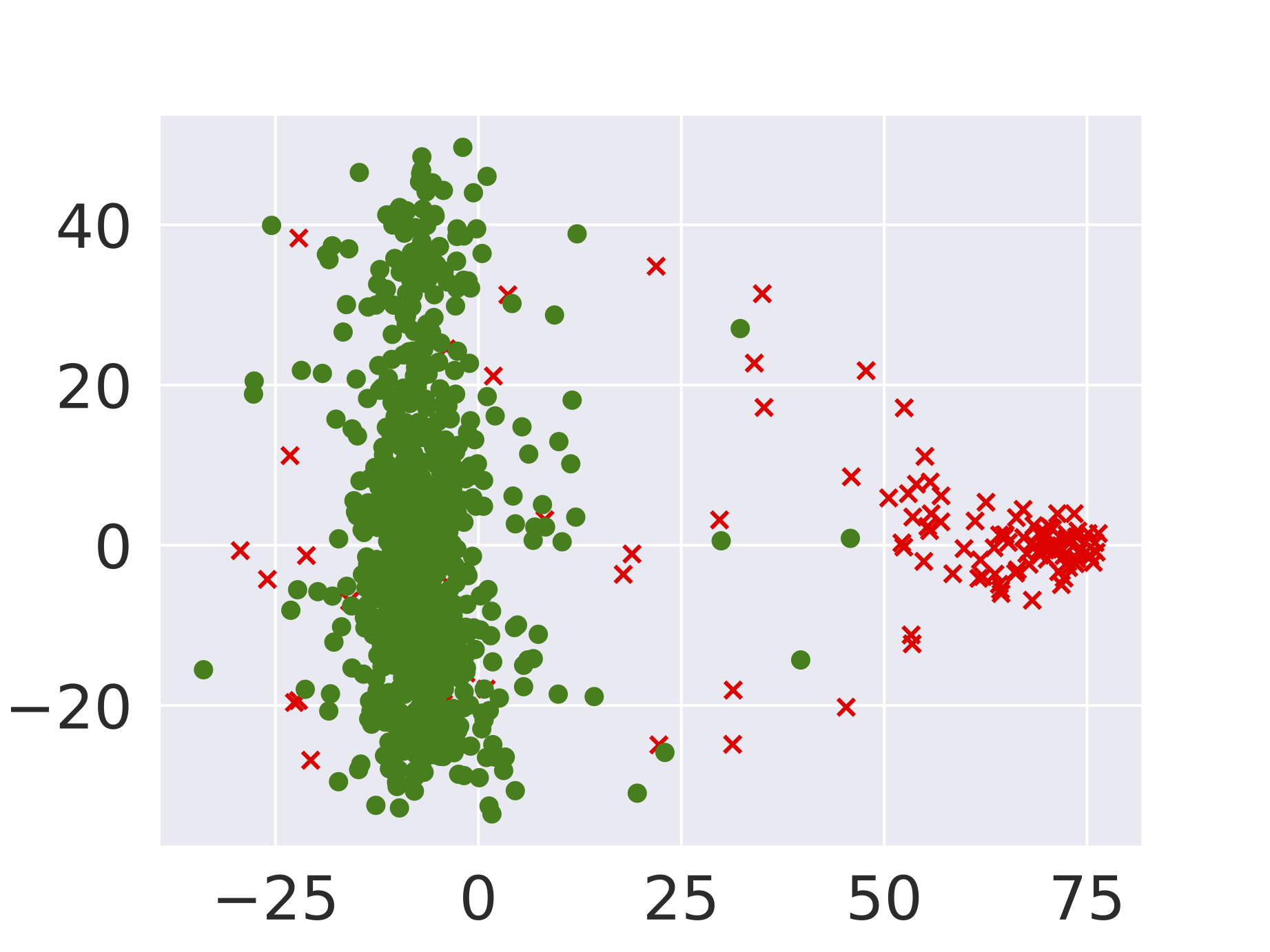}
    \caption{CIFAR-10}
    \label{fig:first3}
\end{subfigure}
\caption{PCA visualization of benign (green) and malicious (red) updates over all communication rounds with $m$=10\%.}
\label{fig:SAfigures}
\vspace{-7mm}
\end{figure}

In this step, we first analyze the final layer updates $\nabla W^i_t$ of the local models trained by each client $i\in \{1,2,3,....,N\}$ at the current round $t$. The honest clients $h \subset i$ primarily focus on increasing the main task accuracy, and train local models on diverse datasets $D_i$ to improve global model performance. Consequently, the benign updates diverge significantly from one another and demonstrate higher variance. In contrast, the poisoned updates show higher similarity, as the malicious clients $m\subset i \setminus h$ share a common goal of increasing the backdoor task accuracy. Therefore, the model updates received from benign and malicious clients follow different gradient distributions. We apply Principal Component Analysis (PCA) on the last layer updates $\nabla W^i_t$ to utilize this difference for detecting poisoned updates. PCA highlights the principal directions of data variation, which facilitates the detection of patterns in the model updates. We observe that the updates tend to form different clusters after reducing the dimensionality to two using PCA. The poisoned updates exhibit higher uniformity while the benign updates exhibit higher variance, as demonstrated in Fig.~\ref{fig:SAfigures}. We focus on the final layer updates because the final layer provides a computationally efficient yet compact representation that captures the essential behavioral differences between honest and malicious clients. Moreover, in targeted poisoning attacks, adversaries primarily manipulate the final classification layer to associate triggers with target labels while preserving benign behavior in earlier layers to evade detection \cite{shafahi2018poison, sun2025partner}. 
\begin{equation}
    Var\;(\nabla W^i_t)_{i \subset h} > Var\;(\nabla W^j_t)_{j \subset m}
\end{equation}

Next, we apply K-Means Clustering to separate the low-dimensional update vectors $\{V_1,V_2,V_3,....,V_N\}$ into two clusters, $C_1$ and $C_2$. As the smaller cluster is not guaranteed to contain malicious updates~\cite{wei2023demystifying}, we further analyze each cluster to avoid misclassifying the updates. We compute the pairwise Cosine Similarity $S$ among the model updates within each cluster as follows. 
\begin{equation}
    S(V_i, V_j) = \frac{V_i \cdot  V_j}{\| V_i\| \| V_j\|}
\end{equation}

The malicious clients demonstrate similar behavior and produce similar updates, resulting in a higher average Cosine Similarity within their cluster. We calculate the average similarity $S_{C_p}$ \eqref{eq8} for each cluster and identify the cluster with the higher similarity value as the malicious cluster. Finally, we perform outlier detection as proposed in \cite{amidan2005data}. We calculate the Euclidean distance of each client update from its assigned cluster centroid. Any update that lies outside the distance threshold is flagged as an outlier and removed. 
\begin{equation}
    S_{C_p} = avg \; { S(V_i, V_j); } \qquad \forall \; {i,j \in C_p, i \neq j}
    \label{eq8}
\end{equation}

\subsection{Temporal Analysis}

We apply the fixed-frequency attack \cite{wang2020attack}, where attackers poison their local datasets only in some communication rounds. By sending poisoned updates less frequently, the attackers become stealthy while still being able to influence the global model~\cite{zhang2022neurotoxin}. In our work, the attackers insert a trigger into the training samples once every third round. 

To identify attackers based on the patterns of their historical updates, we perform per-client temporal analysis. We compute the predicted gradients $R_i^t$ for each client $i$ in the current round $t$ based on its gradients from previous rounds. Specifically, we utilize the exponential moving average of past updates to predict the expected updates in the current round, as given by the following equation. 
\begin{equation}
    R_i^t = \beta R_i^{t-1}+(1-\beta)\Delta W_t
\end{equation}
where $\beta$ is a decay parameter that balances the contributions of the predicted and actual updates. This method allows us to capture the temporal behavior of malicious clients, which is inconsistent over consecutive rounds. Next, we compare the predicted update $R_i^t$ with the actual update $\Delta W^i_t$ sent by client $i$ in the current round $t$. To measure the similarity $\gamma^t_i$ between these two update vectors, we utilize Cosine Similarity \eqref{eq10}. A lower similarity implies that the client’s current behavior is inconsistent with its past behavior, which flags it as a potentially malicious client. 
\begin{equation}
    \gamma^t_i = \frac{R^t_i \cdot \Delta W_t^i}{\|R^t_i\| \,\|\Delta W_t^i\|}
    \label{eq10}
\end{equation}

\begin{figure}
\centering
\begin{subfigure}{0.49\linewidth}
    \includegraphics[width=\linewidth]{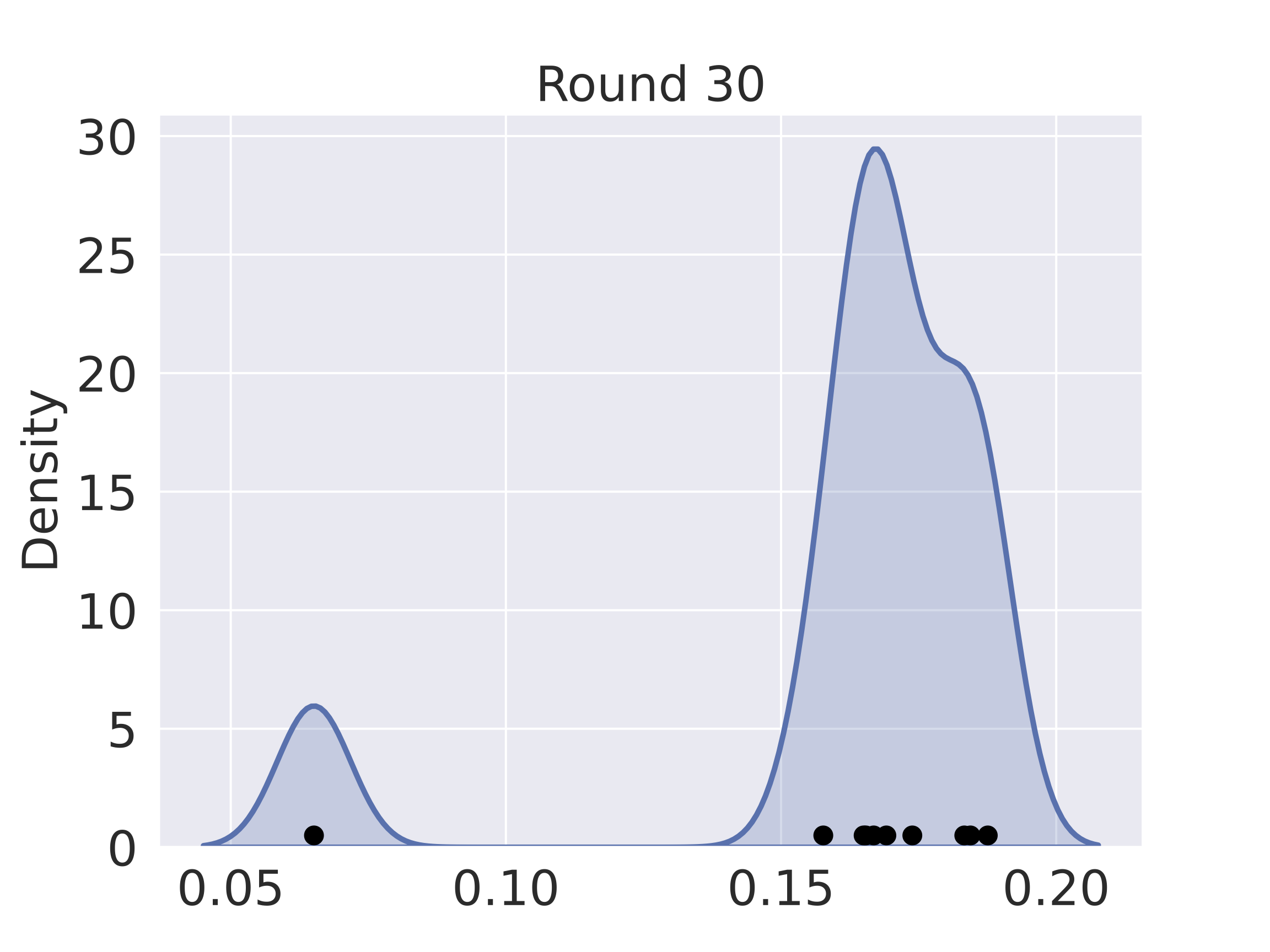}
    \caption{$m$=10\%}
    \label{fig:first4}
\end{subfigure}
\begin{subfigure}{0.49\linewidth}
    \includegraphics[width=\linewidth]{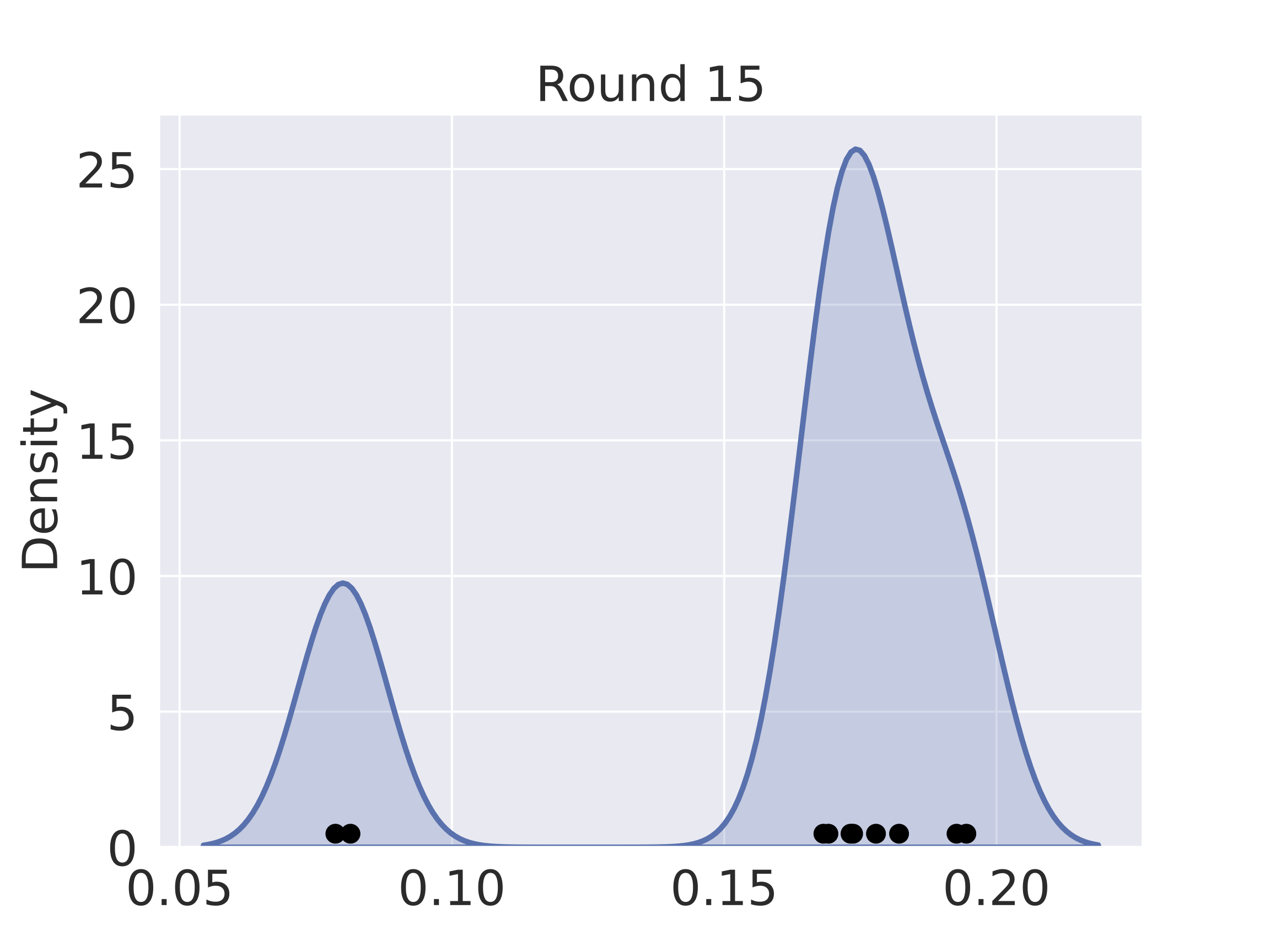}
    \caption{$m$=20\%}
    \label{fig:first5}
\end{subfigure}
\begin{subfigure}{0.49\linewidth}
    \includegraphics[width=\linewidth]{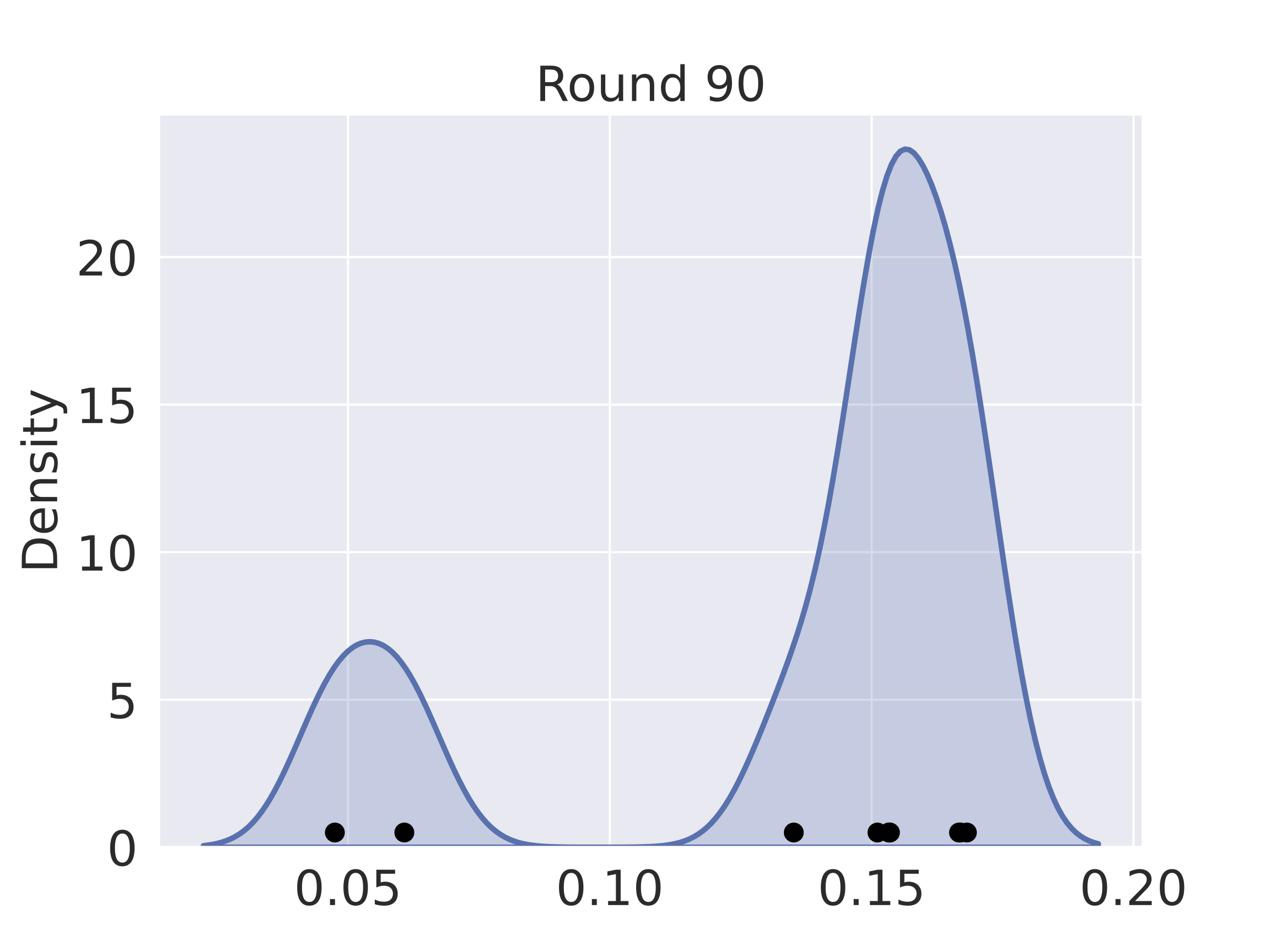}
    \caption{$m$=20\%}
    \label{fig:first6}
\end{subfigure}
\begin{subfigure}{0.49\linewidth}
    \includegraphics[width=\linewidth]{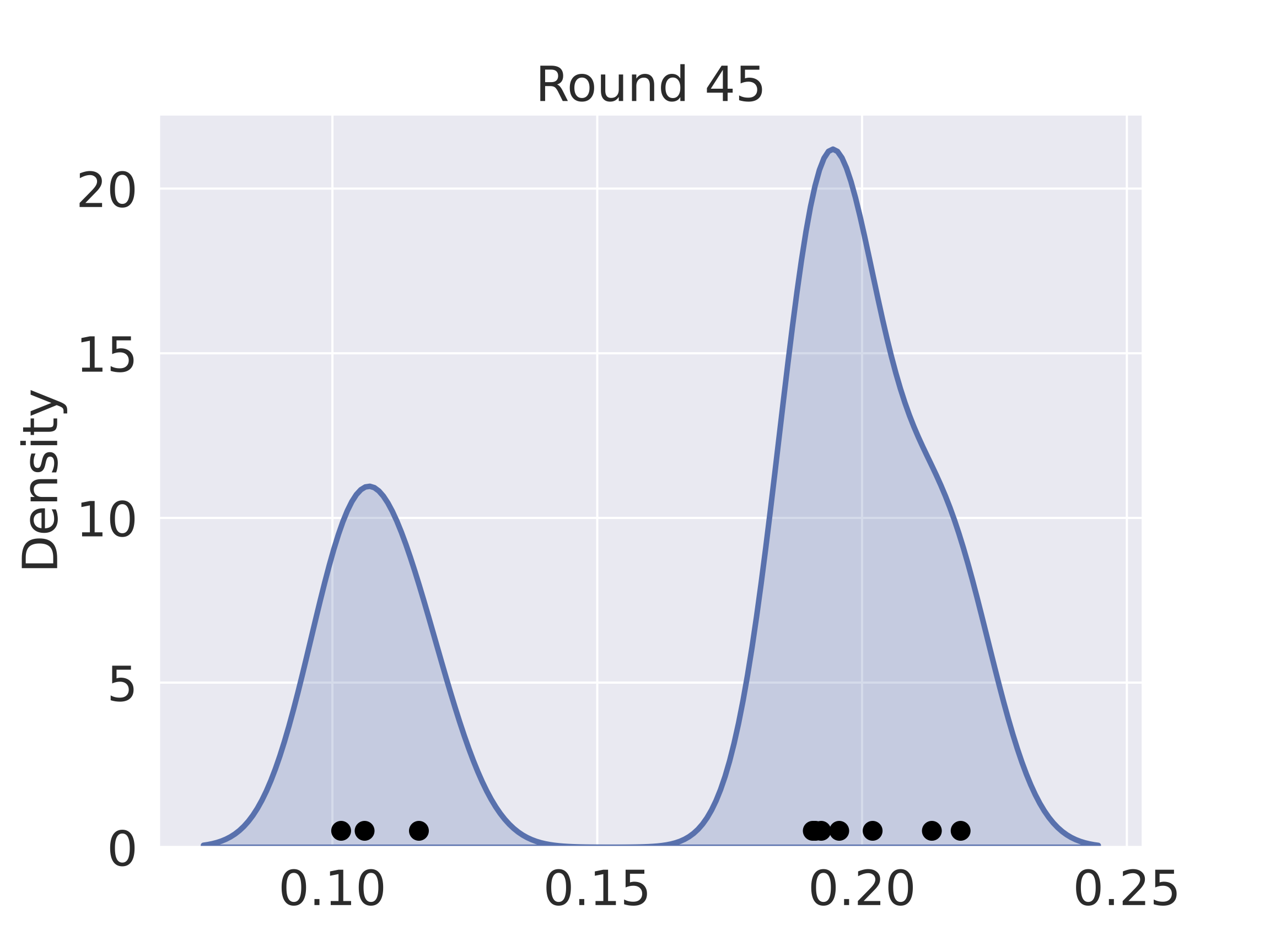}
    \caption{$m$=30\%}
    \label{fig:first7}
\end{subfigure}
        
\caption{Kernel Density Estimation plots at round 30 and 15 for the FMNIST dataset (top), and at round 90 and 45 for the CIFAR-10 dataset (bottom).}
\label{fig:KDEfigures}
\vspace{-6mm}
\end{figure}

Then we apply Kernel Density Estimation (KDE) to the Cosine Similarity values $\{\gamma_1,\gamma_2,\gamma_3,....,\gamma_N\}$. This technique helps us identify the malicious clients based on the density of the Cosine Similarity values. The key idea is that the benign clients' behavior remains consistent over time. Therefore, they form a dense cluster around high similarity scores. In contrast, malicious clients form a separate and smaller cluster as shown in Fig.~\ref{fig:KDEfigures}. Finally, we flag the cluster with fewer clients as the malicious cluster since the number of malicious clients is generally smaller than the number of benign clients. 

\subsection{Robust Aggregation}

In this step, we employ an additional defense mechanism by utilizing adaptive server learning rates. Unlike traditional approaches that maintain a fixed learning rate during aggregation, we dynamically adjust the learning rates across different training rounds. This approach further enhances the robustness of the global model by mitigating the influence of malicious updates. The per-client learning rate tuning strategy is not a practical approach, as malicious clients might manipulate the training process to exploit such a learning rate tuning strategy and enhance their influence on the global model. We adopt a more realistic defense setting, which involves modifying the server learning rate based on the aggregated signs of updates in a particular dimension. 

Let $\triangle W_t^i$ denote the updates received at round $t$ from client $i \in h_t$, where $h_t$ is the set of clients identified as honest. We analyze the signs of updates and aggregate them across clients. Specifically, we compute the sum of signs $M_{t_k}$ for each parameter dimension $k \in \{1,2,3,....,K\}$ as \eqref{eq11} shows. 
\begin{equation}
    M_{t_k}=\sum_{i \in h_t} sign \;(\triangle W_{t_k}^i)
    \label{eq11}
\end{equation}

If the absolute value of the aggregated signs $M_{t_k}$ exceeds a threshold $\delta$, it indicates that the majority of updates are consistent along the dimension $k$. This also indicates that the benign updates are dominant along that dimension, as the number of benign clients is higher than the number of malicious clients in our setting. In this case, the server learning rate $\eta$ is multiplied by 1, which allows the global model to effectively learn from these updates. However, if $|M_{t_k}|$ is less than $\delta$, it implies that the updates do not agree along that dimension. As a result, positive and negative updates tend to cancel each other out. We attribute this inconsistency to the stronger influence of poisoned updates. In such a case, we multiply the learning rate $\eta$ by -1, which minimizes the contribution of malicious updates to the global model. $\delta$ is empirically set to 3 or 4 to achieve a balance between accuracy and attack success rate. 
\begin{equation}
    \eta^t_k = 
        \begin{cases}
        \eta, & if\; |M_{t_k}| > \delta \\
        -\eta, & otherwise
        \end{cases}
\end{equation}

Here, the goal is to mitigate the impact of poisoned updates that evade spatial-temporal detection. By utilizing adaptive learning rates, the influence of these updates is significantly reduced while ensuring that the global model remains strong and robust against adversarial attacks.

\section{Experimental Analysis}

\subsection{Experimental Setup}

\noindent \textbf{Datasets.} We conduct experiments on three benchmark datasets, Fashion-MNIST (FMNIST) \cite{xiao2017fashion}, CIFAR-10 \cite{krizhevsky2009learning}, and CIFAR-100 \cite{krizhevsky2009learning}. The FMNIST dataset consists of 60,000 training and 10,000 testing images, and the CIFAR-10 and CIFAR-100 datasets consist of 50,000 training and 10,000 testing images. Both FMNIST and CIFAR-10 datasets consist of 10 classes, and CIFAR-100 consists of 100 classes.

\noindent \textbf{Model Architectures.} For the FMNIST dataset, we use a Convolutional Neural Network (CNN) consisting of two convolutional layers, one max pooling layer, and two fully connected layers with dropout. For CIFAR-10, we use a CNN with three convolutional layers, one max pooling layer, and three fully connected layers with dropout. Finally, we use a ResNet-18 architecture to train the CIFAR-100 dataset. 

\noindent \textbf{Attack Method.} We consider 20 clients in our setting, with 50\% of the clients randomly chosen in each round to participate in the FL process. The attackers insert a plus pattern as a trigger (Fig.~\ref{fig:figures}) into all samples of a specific class (base class) to poison the local datasets following the BadNets \cite{gu2017badnets} attack. Then they perform the following label flipping: class sandal $\rightarrow$ sneaker (FMNIST), class dog $\rightarrow$ horse (CIFAR-10), and class bed $\rightarrow$ beetle (CIFAR-100).

\noindent \textbf{Hyperparameters.} We run the FL training for 30, 100, and 100 rounds on FMNIST, CIFAR-10, and CIFAR-100 datasets, respectively. The learning rate for local training is set to 0.1, and the server learning rate is initially set to 1.0. The batch size is set to 256 for FMNIST and CIFAR-10, and 128 for CIFAR-100. The number of epochs is 5 for the FMNIST dataset and 2 for the CIFAR-10 and CIFAR-100 datasets. The FMNIST and CIFAR-10 experiments are performed in an IID setting, while the CIFAR-100 experiments are performed in a non-IID setting with Dirichlet parameter $\alpha$ = 0.7. All experiments are conducted using the PyTorch framework.

\noindent \textbf{Evaluation Metrics.} We use three evaluation metrics: main task accuracy (MA), base class accuracy (BA), and attack success rate (ASR), to assess the performance of the global model in our adversarial setting. MA and BA are measured on a clean validation set (without trigger), while ASR is measured on a poisoned validation set containing only poisoned samples from the base class. The goal of our defense is to maximize MA and BA while minimizing ASR.

\noindent \textbf{a) Main Task Accuracy:} the overall accuracy, which indicates how accurately the model classifies clean samples.

\noindent \textbf{b) Base Class Accuracy:} the accuracy on the base class, which indicates how well the model classifies clean samples from the base class.

\noindent  \textbf{c) Attack Success Rate:} the percentage of base class samples that are classified into the attacker-specified target class in the presence of a trigger.

\subsection{Effectiveness of \projectname{}}

\begin{table}[!t]
\centering
\caption{Main task accuracy, base class accuracy, and attack success rate with different percentages of malicious clients. $\uparrow$ denotes better performance with higher values, and $\downarrow$ denotes better performance with lower values.}
\scalebox{0.965}{
\begin{tabular}{|c|c|c|c|c|}
\hline
Dataset                  & \begin{tabular}[c]{@{}c@{}}\% of Malicious\\ Clients\end{tabular} & \begin{tabular}[c]{@{}c@{}}MA (\%)($\uparrow$)\end{tabular} & \begin{tabular}[c]{@{}c@{}}BA (\%)($\uparrow$)\end{tabular} & \begin{tabular}[c]{@{}c@{}}ASR (\%)($\downarrow$)\end{tabular} \\ \hline
\multirow{3}{*}{FMNIST}  & 10                                                               & \textbf{91.49}                                                              &\textbf{97.50}                                                               & \textbf{2.30}                                                              \\ \cline{2-5} 
                         & 20                                                               & 90.94                                                              & 97.10                                                               & 2.90                                                              \\ \cline{2-5} 
                         & 30                                                               & 89.96                                                              & 96.80                                                               & 3.60                                                              \\ \hline
\multirow{3}{*}{CIFAR-10} & 10                                                               & \textbf{74.63}                                                              & 57.70                                                               & \textbf{3.20}                                                              \\ \cline{2-5} 
                         & 20                                                               & 74.61                                                              & \textbf{59.40}                                                               & 3.90                                                              \\ \cline{2-5} 
                         & 30                                                               & 72.58                                                              & 52.60                                                               & 4.10                                                              \\ \hline
\end{tabular}
} 
\label{table1}
\vspace{-5mm}
\end{table}

\begin{table*}[h]
\centering
\caption{Effectiveness of our proposed defense compared to existing defenses. $\uparrow$ denotes better performance with higher values, and $\downarrow$ denotes better performance with lower values.}
\begin{tabular}{|c|cc|cc|cc|}
\hline
             & \multicolumn{2}{c|}{FMNIST} & \multicolumn{2}{c|}{CIFAR-10} & \multicolumn{2}{c|}{CIFAR-100} \\ \hline
Defense      & \multicolumn{1}{c|}{\begin{tabular}[c]{@{}c@{}}Main Task\\ Accuracy (\%)($\uparrow$)\end{tabular}} & \begin{tabular}[c]{@{}c@{}}Attack Success\\ Rate (\%)($\downarrow$)\end{tabular} & \multicolumn{1}{c|}{\begin{tabular}[c]{@{}c@{}}Main Task\\ Accuracy (\%)($\uparrow$)\end{tabular}} & \begin{tabular}[c]{@{}c@{}}Attack Success\\ Rate (\%)($\downarrow$)\end{tabular} & \multicolumn{1}{c|}{\begin{tabular}[c]{@{}c@{}}Main Task\\ Accuracy (\%)($\uparrow$)\end{tabular}} & \begin{tabular}[c]{@{}c@{}}Attack Success\\ Rate (\%)($\downarrow$)\end{tabular} \\ \hline
No Attack    & \multicolumn{1}{c|}{92.52} & N/A   & \multicolumn{1}{c|}{78.63} & N/A   & \multicolumn{1}{c|}{59.31} & N/A \\ \hline
FedAvg \cite{mcmahan2017communication} & \multicolumn{1}{c|}{92.33} & 100.0 & \multicolumn{1}{c|}{78.11} & 77.50 & \multicolumn{1}{c|}{\textbf{59.25}} & 52.00 \\ \hline
RFA \cite{pillutla2022robust}          & \multicolumn{1}{c|}{\textbf{92.41}} & 100.0 & \multicolumn{1}{c|}{\textbf{78.40}} & 78.80 & \multicolumn{1}{c|}{55.68} & 47.00 \\ \hline
Trimmed Mean \cite{yin2018byzantine}  & \multicolumn{1}{c|}{92.29} & 99.50 & \multicolumn{1}{c|}{77.02} & 68.90 & \multicolumn{1}{c|}{57.90} & 44.00 \\ \hline
Median \cite{yin2018byzantine}        & \multicolumn{1}{c|}{91.31} & 77.40 & \multicolumn{1}{c|}{72.38} & 66.00 & \multicolumn{1}{c|}{42.81} & 34.00 \\ \hline
RLR \cite{ozdayi2021defending}        & \multicolumn{1}{c|}{90.98} & 48.60 & \multicolumn{1}{c|}{74.68} & 43.60 & \multicolumn{1}{c|}{51.73} & 21.00 \\ \hline
FoolsGold \cite{fung2018mitigating}        & \multicolumn{1}{c|}{91.88} & 39.50 & \multicolumn{1}{c|}{74.98} & 71.30 & \multicolumn{1}{c|}{57.83} & 43.00 \\ \hline
STAR-FL                               & \multicolumn{1}{c|}{89.96} & \textbf{3.60} & \multicolumn{1}{c|}{72.58} & \textbf{4.10} & \multicolumn{1}{c|}{53.62} & \textbf{7.00} \\ \hline
\end{tabular}
\label{table2}
\vspace{-5mm}
\end{table*}

To evaluate the effectiveness of the proposed STAR-FL framework, we conducted experiments under varying proportions of malicious clients ($m$ = 10\%/20\%/30\%). Table~\ref{table1} presents the experimental results under different attack fractions. On the FMNIST dataset, STAR-FL demonstrated strong resilience against adversarial manipulation. With 10\% malicious clients, the framework achieved a main task accuracy of 91.49\%, maintaining a high base class accuracy of 97.50\% while limiting the ASR to only 2.30\%. As the percentage of malicious clients increased to 20\% and 30\%, there was a gradual decline in overall accuracy. However, the degradation remained moderate. Even with 30\% malicious client participation, the framework sustained a main task accuracy of 89.96\% with an ASR of 3.60\%. This highlights the robustness of STAR-FL in mitigating adversarial influence while maintaining main task performance.

The experiments on the CIFAR-10 dataset present a more challenging scenario due to the increased complexity of the data. With 10\% malicious clients, the framework obtained a main task accuracy of 74.63\% and a base class accuracy of 57.70\%, with the ASR remaining as low as 3.20\%. The main task accuracy was preserved when the proportion of malicious clients increased. However, a sharper decline in base class accuracy was observed. With 30\% malicious clients, the base class accuracy dropped to 52.60\%, while the ASR rose slightly to 4.10\%. These results indicate that while STAR-FL is effective in constraining the success rate of backdoor attacks, it incurs some trade-offs in maintaining accuracy on complex datasets.

In summary, the results demonstrate that STAR-FL achieves a balance between robustness and accuracy across different attack fractions. The framework consistently restricted the ASR to below 5\% for both datasets. This suggests that STAR-FL provides a significant advantage in isolating malicious updates, thereby enhancing the security of federated learning systems in adversarial environments. We further evaluate the effectiveness of STAR-FL against advanced attacks. The results shown in Section \ref{sec:diffattack} demonstrate that STAR-FL provides a resilient and flexible defense against diverse attack strategies.

\subsection{Comparison to Existing Defenses}
We compare STAR-FL with RFA \cite{pillutla2022robust}, Trimmed Mean \cite{yin2018byzantine}, Median \cite{yin2018byzantine}, RLR \cite{ozdayi2021defending}, and FoolsGold \cite{fung2018mitigating}. Table~\ref{table2} shows the experimental results with 30\% malicious clients. The CIFAR-100 experiments are conducted in a continuous attack~\cite{xie2019dba} setting where attackers poison datasets in every round. Moreover, they scale their updates to amplify the attack. For all non-IID experiments, we determine the optimal number of clusters in spatial analysis using the Gap Statistics \cite{tibshirani2001estimating} method instead of fixing the setup to two clusters, and we omit the outlier-rejection step used in the IID experiments. These changes ensure that under non-IID settings, the framework does not mistakenly exclude honest updates while isolating malicious clients.

On FMNIST, CIFAR-10, and CIFAR-100 datasets, FedAvg achieves high main task accuracy but completely fails as a defense, with an ASR of 100\% on FMNIST, 77.50\% on CIFAR-10, and 52\% on CIFAR-100. RFA aggregates updates using the geometric median to reduce the effect of outliers. It preserves accuracy almost as well as FedAvg. However, it offers low robustness, with ASR values of 100\%, 78.80\%, and 47\% on FMNIST, CIFAR-10, and CIFAR-100, which indicates that geometric median based aggregation is ineffective against carefully designed malicious updates. Trimmed Mean discards the largest and smallest coordinate updates before averaging. It shows marginal improvement compared to FedAvg and RFA in terms of ASR. However, the effect remains negligible as accuracy is nearly unchanged. 

Median is another coordinate-wise aggregation method that selects the middle updates at each dimension. It provides stronger robustness by lowering ASR to 77.40\% on FMNIST, 66\% on CIFAR-10, and 34\% on CIFAR-100. However, it achieves low accuracy, especially on CIFAR-100, which reflects how discarding benign information can degrade model performance. RLR is an adaptive defense that adjusts the server learning rate according to client update consistency. It achieves stronger defense results by lowering ASR to 48.60\%, 43.60\%, and 21\% on FMNIST, CIFAR-10, and CIFAR-100. However, RLR depends entirely on sign consistency of updates, which makes it vulnerable when malicious updates are crafted to mimic benign updates. FoolsGold partially eliminates the effect of poisoned updates compared to other baselines. On FMNIST, it achieves high main task accuracy (91.88\%) and a relatively low ASR (39.50\%), which indicates that it can effectively preserve main task accuracy while reducing ASR on simpler datasets. However, on CIFAR-10 and CIFAR-100, FoolsGold exhibits higher ASRs of 71.30\% and 43\%, which suggests that the attack can still succeed in more complex feature spaces.

In contrast, STAR-FL demonstrates higher robustness, reducing ASR to 3.60\%, 4.10\%, and 7\% on FMNIST, CIFAR-10, and CIFAR-100. Although STAR-FL incurs a modest reduction in accuracy, its drastic improvement in ASR clearly outweighs this trade-off. By integrating spatial–temporal filtering with robust aggregation, STAR-FL establishes a more reliable and stronger defense framework for federated learning.

\subsection{Ablation Studies}

\begin{table}[t!]
\centering
\caption{Main task accuracy, base class accuracy, and attack success rate under different combinations of spatial and temporal filtering. $\uparrow$ denotes better performance with higher values, and $\downarrow$ denotes better performance with lower values.}
\begin{tabular}{|c|c|c|c|c|}
\hline
Dataset                  & Defense          & \begin{tabular}[c]{@{}c@{}}MA (\%)($\uparrow$)\end{tabular} & \begin{tabular}[c]{@{}c@{}}BA (\%)($\uparrow$)\end{tabular} & \begin{tabular}[c]{@{}c@{}}ASR (\%)($\downarrow$)\end{tabular} \\ \hline
\multirow{3}{*}{FMNIST}  & Spatial          & 92.10                                                              & 98.29                                                               & 2.00                                                                \\ \cline{2-5} 
                         & Temporal         & 92.18                                                              & \textbf{98.79}                                                               & \textbf{1.90}                                                                \\ \cline{2-5} 
                         & Spat.+Temp. & \textbf{92.19}                                                              & 98.19                                                               & 2.60                                                                \\ \hline
\multirow{3}{*}{CIFAR-10} & Spatial          & 78.22                                                              & \textbf{69.20}                                                               & 4.60                                                                \\ \cline{2-5} 
                         & Temporal         & \textbf{78.26}                                                              & 67.69                                                               & 6.50                                                                \\ \cline{2-5} 
                         & Spat.+Temp. & 78.01                                                              & 63.80                                                               & \textbf{4.20}                                                                \\ \hline
\end{tabular}
\label{table3}
\vspace{-1mm}
\end{table}
\begin{table}[t!]
\centering
\caption{Main task accuracy, base class accuracy, and attack success rate under different combinations of spatial-temporal filtering and robust aggregation. $\uparrow$ denotes better performance with higher values, and $\downarrow$ denotes better performance with lower values.}
\begin{tabular}{|c|c|c|c|c|}
\hline
Dataset                  & Defense            & \begin{tabular}[c]{@{}c@{}}MA (\%)($\uparrow$)\end{tabular} & \begin{tabular}[c]{@{}c@{}}BA (\%)($\uparrow$)\end{tabular} & \begin{tabular}[c]{@{}c@{}}ASR (\%)($\downarrow$)\end{tabular} \\ \hline
\multirow{3}{*}{FMNIST}  & Spat.+Robust     & \textbf{91.34}                                                              & 97.60                                                               & 2.80                                                                \\ \cline{2-5} 
                         & Temp.+Robust    & 91.29                                                              & \textbf{97.89}                                                               & 2.70                                                                \\ \cline{2-5} 
                         & Robust & 91.32                                                              & 96.79                                                               & \textbf{1.30}                                                                \\ \hline
\multirow{3}{*}{CIFAR-10} & Spat.+Robust     & 73.95                                                              & 53.10                                                               & \textbf{4.60}                                                                \\ \cline{2-5} 
                         & Temp.+Robust    & 73.97                                                              & \textbf{58.99}                                                               & 6.10                                                                \\ \cline{2-5} 
                         & Robust & \textbf{75.32}                                                              & 57.40                                                               & 4.70                                                                \\ \hline
\end{tabular}
\label{table4}
\vspace{-5mm}
\end{table}

\textbf{Effectiveness of Spatial-Temporal Analysis.} We examined the effectiveness of spatial and temporal filtering, both individually and in combination. These experiments
were conducted with 10\% malicious clients. Table~\ref{table3} highlights the results for these three defense settings.

On the FMNIST dataset, spatial and temporal filtering individually achieved comparable performance, with main task accuracies of 92.10\% and 92.18\%, respectively. Temporal filtering offered the lowest ASR (1.90\%) and the highest base class accuracy (98.79\%), demonstrating its efficacy in mitigating adversarial behavior. The combined spatial+temporal defense achieved the highest main task accuracy (92.19\%) with a marginal increase in ASR. These results indicate that both defenses are highly effective on simpler datasets, with temporal filtering providing a small advantage in terms of robustness. On the more complex CIFAR-10 dataset, spatial filtering outperformed temporal filtering, achieving a lower ASR and higher base class accuracy. The combined spatial+temporal defense reduced the ASR further to 4.20\%, though at the cost of a noticeable drop in base class accuracy. This trade-off highlights the difficulty of distinguishing malicious updates from benign updates in high-dimensional settings, where stricter filtering improves robustness but reduces accuracy.

Overall, the results confirm that STAR-FL effectively constrains backdoor attack success, consistently keeping ASR below 7\%. While temporal filtering is slightly more effective on simpler data, the combined approach provides stronger robustness in complex scenarios, validating the effectiveness of spatial and temporal filtering components of STAR-FL.

\textbf{Effectiveness of Robust Aggregation.} We further examined the performance of STAR-FL by incorporating robust aggregation with spatial and temporal filtering. Table~\ref{table4} highlights the results for these defense settings. On the FMNIST dataset, all defense combinations maintained stable performance, with main task accuracy around 91\%. Spatial+robust and temporal+robust achieved comparable results, with temporal+robust slightly outperforming in terms of base class accuracy (97.89\%) and ASR (2.70\%). Robust aggregation alone reduced the ASR most effectively to 1.30\%, though at the cost of a lower base class accuracy. These findings indicate that the integration of robust aggregation suppresses ASR, though sometimes at the cost of accuracy. 

On the CIFAR-10 dataset, spatial+robust and temporal+robust yielded main task accuracies of 73.95\% and 73.97\%, respectively, with temporal+robust achieving the highest base class accuracy (58.99\%) and the highest ASR (6.10\%). Robust aggregation alone provided the best balance, achieving the highest main task accuracy (75.32\%) and keeping the ASR low (4.70\%), while also maintaining a competitive base class accuracy (57.40\%). Therefore, robust aggregation demonstrates a favorable trade-off by maintaining high accuracy while keeping ASR under control.

Overall, these findings highlight that STAR-FL benefits from the complementary strengths of spatial-temporal filtering and robust aggregation, while offering flexible defense depending on dataset complexity. On simpler datasets such as FMNIST, spatial-temporal filtering provides limited benefit and robust aggregation alone outperforms STAR-FL in terms of ASR. In contrast, the full framework provides greater benefits on complex datasets such as CIFAR-10.

\begin{table*}[!h]
\centering
\caption{Main task accuracy and attack success rate under various data heterogeneity. $\uparrow$ denotes better performance with higher values, and $\downarrow$ denotes better performance with lower values.}
\label{table6}
\begin{tabular}{|c|c|cc|cc|}
\hline
\multirow{3}{*}{$\alpha$} & \multirow{2}{*}{\begin{tabular}[c]{@{}c@{}}\% of \\ malicious\\ clients\end{tabular}} & \multicolumn{2}{c|}{FMNIST}                                                                                                                                     & \multicolumn{2}{c|}{CIFAR-10}                                                                                                                                    \\ \cline{3-6} 
                       &                                                                                       & \multicolumn{1}{c|}{\begin{tabular}[c]{@{}c@{}}Main Task\\ Accuracy (\%)($\uparrow$)\end{tabular}} & \begin{tabular}[c]{@{}c@{}}Attack Success\\ Rate (\%)($\downarrow$)\end{tabular} & \multicolumn{1}{c|}{\begin{tabular}[c]{@{}c@{}}Main Task\\ Accuracy (\%)($\uparrow$)\end{tabular}} & \begin{tabular}[c]{@{}c@{}}Attack Success\\ Rate (\%)($\downarrow$)\end{tabular} \\ \hline
\multirow{3}{*}{0.5}   & 10                                                                                    & \multicolumn{1}{c|}{79.32}                                                               & 2.50                                                                 & \multicolumn{1}{c|}{46.15}                                                               & \textbf{3.20}                                                                 \\ \cline{2-6} 
                       & 20                                                                                    & \multicolumn{1}{c|}{78.16}                                                               & 6.10                                                                 & \multicolumn{1}{c|}{42.19}                                                               & 7.80                                                                 \\ \cline{2-6} 
                       & 30                                                                                    & \multicolumn{1}{c|}{79.24}                                                               & 20.10                                                                & \multicolumn{1}{c|}{42.68}                                                               & 13.00                                                                \\ \hline
\multirow{3}{*}{1.0}   & 10                                                                                    & \multicolumn{1}{c|}{\textbf{88.91}}                                                               & \textbf{2.10}                                                                 & \multicolumn{1}{c|}{\textbf{58.05}}                                                               & 8.40                                                                 \\ \cline{2-6} 
                       & 20                                                                                    & \multicolumn{1}{c|}{86.56}                                                               & 4.10                                                                 & \multicolumn{1}{c|}{52.43}                                                               & 11.60                                                                \\ \cline{2-6} 
                       & 30                                                                                    & \multicolumn{1}{c|}{88.24}                                                               & 9.10                                                                 & \multicolumn{1}{c|}{56.33}                                                               & 10.60                                                                \\ \hline
\end{tabular}
\vspace{-3mm}
\end{table*} 

\textbf{Effectiveness on Non-IID Data.} To further evaluate the performance of STAR-FL in non-IID settings, we experimented with 30\% malicious clients and varied the parameter $\alpha$ of the Dirichlet partitioning to control data heterogeneity. In a Dirichlet distribution, lower $\alpha$ means more skewed (non-IID) data, while higher $\alpha$ approaches an IID distribution. The results in Table~\ref{table6} show that greater heterogeneity degrades accuracy and increases ASR. At $\alpha$=0.5, the accuracies drop and ASRs rise substantially compared to $\alpha$=1.0. For example, with 30\% malicious clients on FMNIST, the main task accuracy is  $\approx$79\% at $\alpha$=0.5 and $\approx$88\% at $\alpha$=1.0, and the ASR jumps from $\approx$9\% ($\alpha$=1.0) to $\approx$20\% ($\alpha$=0.5). On CIFAR-10, the accuracy improves from $\approx$43\% to $\approx$56\% when moving from $\alpha$=0.5 to 1.0, while ASR remains moderate ($\approx$10–13\%). These results indicate that our defense is still effective when client data distributions are imbalanced. 

In summary, increased heterogeneity makes it slightly harder for STAR-FL to separate malicious updates from benign ones, since malicious updates also become highly diverse. However, the results in non-IID settings show that STAR-FL can maintain robustness even under data heterogeneity through spatial-temporal filtering and robust aggregation.

\label{sec:diffattack}
\textbf{Effectiveness on Different Attacks.} We measured the performance of \projectname{} against different types of attacks: Model Replacement Attack (MRA) \cite{bagdasaryan2020backdoor}, Distributed Backdoor Attack (DBA) \cite{xie2019dba}, a combined MRA+DBA attack, and an adaptive attack. These experiments were conducted with 30\% malicious clients to simulate stronger attack scenarios. For the MRA attack, malicious updates are multiplied by a scaling factor before sending to the server. For the DBA attack, the backdoor trigger is distributed among three attackers, who jointly form a pattern as the trigger. Table~\ref{table5} summarizes the experimental results under these attacks.

\begin{table}[!t]
\centering
\caption{Main task accuracy, base class accuracy, and attack success rate under different attacks. $\uparrow$ denotes better performance with higher values, and $\downarrow$ denotes better performance with lower values.}
\begin{tabular}{|c|c|c|c|c|}
\hline
Attack                                                                                                & Dataset & \begin{tabular}[c]{@{}c@{}}MA (\%)($\uparrow$)\end{tabular} & \begin{tabular}[c]{@{}c@{}}BA (\%)($\uparrow$)\end{tabular} & \begin{tabular}[c]{@{}c@{}}ASR (\%)($\downarrow$)\end{tabular} \\ \hline
\multirow{2}{*}{\begin{tabular}[c]{@{}c@{}}MRA\end{tabular}}                   & FMNIST  & 89.32                                                              & 96.10                                                               & 3.60                                                                \\ \cline{2-5} 
                                                                                                      & CIFAR-10 & 74.40                                                              & 58.89                                                               & \textbf{3.30}                                                                \\ \hline
\multirow{2}{*}{\begin{tabular}[c]{@{}c@{}}DBA\end{tabular}}                & FMNIST  & \textbf{89.62}                                                              & \textbf{96.70}                                                               & \textbf{3.20}                                                                \\ \cline{2-5} 
                                                                                                      & CIFAR-10 & 73.43                                                              & 55.10                                                               & 5.10                                                                \\ \hline
\multirow{2}{*}{\begin{tabular}[c]{@{}c@{}}MRA+DBA \end{tabular}} & FMNIST  & 89.39                                                              & 95.49                                                               & 3.63                                                                \\ \cline{2-5} 
                                                                                                      & CIFAR-10 & 75.54                                                              & 58.79                                                               & 4.30                                                                \\ \hline \multirow{2}{*}{\begin{tabular}[c]{@{}c@{}}Adaptive\\ Attack \end{tabular}} & FMNIST  & 89.48                                                              & 92.80                                                               & 3.40                                                                \\ \cline{2-5} 
                                                                                                      & CIFAR-10 & \textbf{75.86}                                                              & \textbf{62.80}                                                               & 4.30                                                                \\ \hline
\end{tabular}
\vspace{-5mm}
\label{table5}
\end{table}

On the FMNIST dataset, STAR-FL demonstrated consistent resilience against all types of attacks. Under the MRA attack, STAR-FL maintained a main task accuracy of 89.32\% and base class accuracy of 96.10\%, while keeping the ASR relatively low at 3.60\%. Under the DBA attack, ASR was reduced to 3.20\%, with a slight increase in main task accuracy and base class accuracy. When the MRA and DBA attacks were combined, ASR slightly increased to 3.63\%, reflecting the cumulative challenge posed by concurrent attack strategies. The base class accuracy also decreased moderately to 95.49\%. These results indicate that spatial-temporal filtering and robust aggregation effectively mitigate backdoor threats, although complex simultaneous attacks may have a measurable impact on model performance.

On the CIFAR-10 dataset, the impact of the attacks was more noticeable. The MRA attack resulted in a main task accuracy of 74.40\% and base class accuracy of 58.89\%, with an ASR of 3.30\%. The DBA attack resulted in a main task accuracy of 73.43\% and an ASR of 5.10\%, which implies that DBA is more effective in high-dimensional datasets. Interestingly, the MRA+DBA caused the main task accuracy to increase to 75.54\%, while keeping the ASR relatively low at 4.30\%. This indicates that the combination of different attacks can produce non-trivial effects on model performance.

The robustness of STAR-FL is further evaluated under an adaptive attack where adversaries modify the attack to bypass detection. The adaptive attack consists of three steps: 1) Gaussian noise is added to poisoned updates, 2) the attack is performed in every communication round, and 3) multiple triggers are injected by each malicious client. Despite these attacks, STAR-FL maintains strong robustness, achieving 89.48\% accuracy with 3.40\% ASR on FMNIST and 75.86\% accuracy with 4.30\% ASR on CIFAR-10. These results demonstrate that STAR-FL can effectively detect adaptive adversaries even when they attempt to bypass the defense.

In summary, while sophisticated attacks may slightly affect accuracy and ASR, the combination of spatial-temporal filtering and robust aggregation enables reliable suppression of attacks while maintaining stable model performance.

\subsection{Computation and Communication Overhead of \projectname{}}
\label{sec:overhead}

\begin{figure}
\centering
\includegraphics[width=\linewidth]{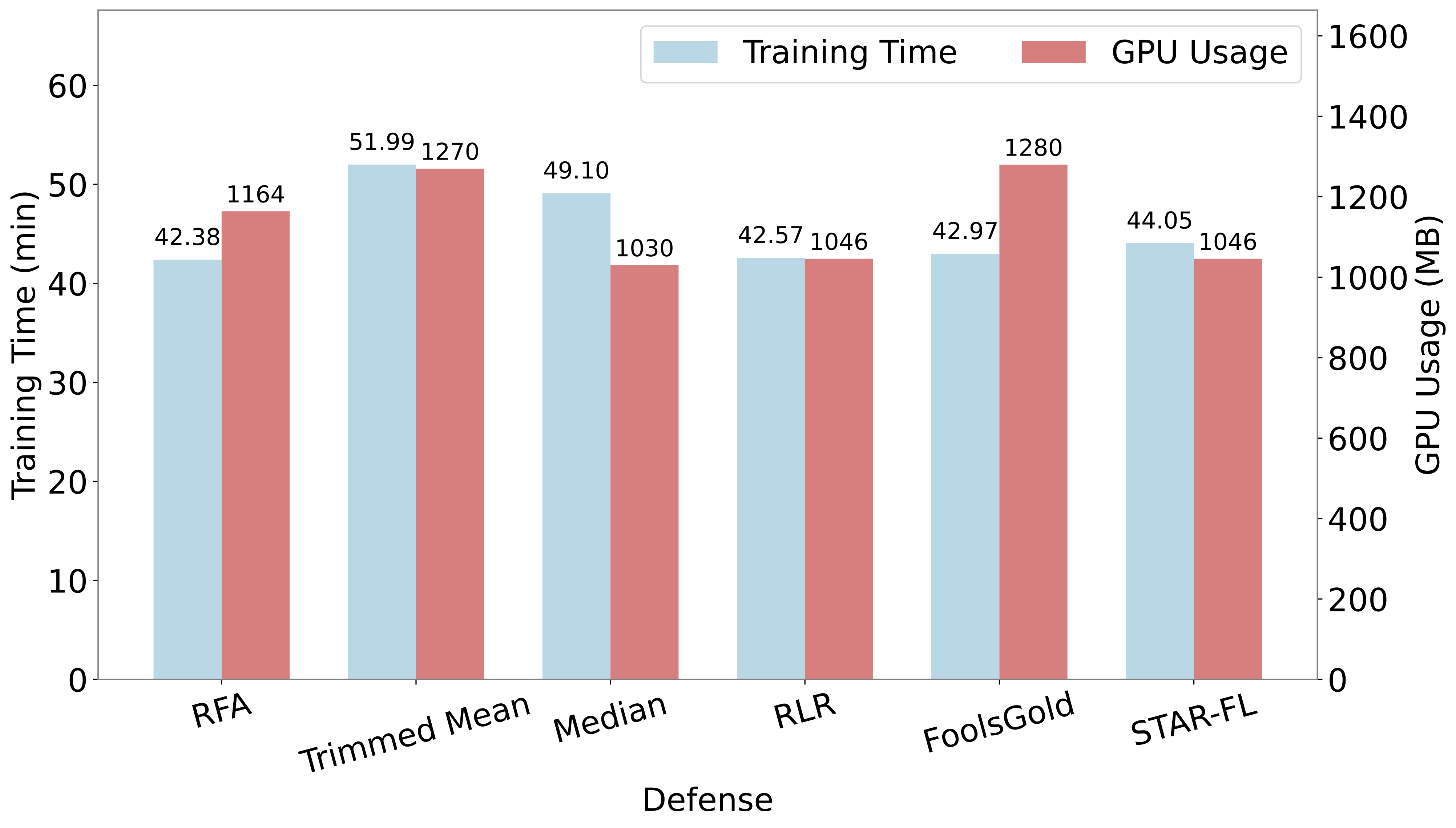}
    \caption{Training time and GPU usage overhead of STAR-FL compared to baseline methods on CIFAR-10.}
    \label{overhead}
    \vspace{-6mm}
\end{figure}

We evaluate the computational overhead of STAR-FL to assess its feasibility. Specifically, we measure the training time and GPU memory usage on CIFAR-10 as shown in Fig.~\ref{overhead}. The spatial-temporal analysis and learning rate adjustments introduce some overhead at the server. STAR-FL requires $\approx$44 minutes and $\approx$1 GB, which is close to the training time and GPU usage of RLR. Trimmed Mean and FoolsGold incur the highest training time and GPU usage respectively. The coordinate-wise sorting operations in Trimmed Mean and the history‑based similarity checking in FoolsGold introduce significant computational complexity. Overall, STAR-FL is 10-15\% faster than Trimmed Mean and Median. Regarding GPU usage, STAR-FL utilizes 10-18\% less memory than RFA, Trimmed Mean, and FoolsGold. The results indicate that while STAR-FL performs additional clustering and learning rate tuning, these steps are efficiently implemented and do not substantially increase computational costs.

STAR-FL does not introduce any additional communication overhead. In each round, clients transmit model updates of size $s$ (bytes), and the server sends the aggregated global model back to the clients. Let $c$ denote the number of participating clients per round and $r$ the number of rounds. The total communication cost is $2\cdot s\cdot c\cdot r$ bytes. The spatial and temporal analyses are performed entirely on the received updates at the server side and do not require any additional information from clients. Consequently, the per-round communication cost of STAR-FL remains identical to FedAvg.

\section{Conclusion}
In this paper, we present \projectname{}, a novel defense framework that enhances the security and robustness of federated learning systems against backdoor attacks. 
First, we utilize spatial-temporal analysis to jointly examine the similarity among client updates and their historical consistency across training rounds, distinguishing benign from malicious behaviors. 
Second, we develop a clustering-based detection mechanism that identifies malicious updates based on their spatial characteristics while leveraging temporal patterns of client updates to detect abnormal behaviors that evolve over time. 
Third, we integrate a robust aggregation strategy that adaptively adjusts the learning rates for client updates to mitigate the influence of adversarial contributions that evade detection. 
By combining spatial-temporal analysis with robust aggregation, \projectname{} provides a unified defense framework that improves robustness against backdoor attacks. Extensive experiments demonstrate that \projectname{} outperforms state-of-the-art defense methods, while maintaining high model utility and significantly reducing attack success rates. 

\bibliographystyle{ieeetr}
\bibliography{main}

\end{document}